\documentclass[aps,pra,twocolumn,floatfix]{revtex4}
\usepackage{graphicx}
\usepackage{dcolumn}
\usepackage{bm}
\usepackage{hyperref}
\usepackage{amssymb}
\usepackage{amsmath}
\usepackage{amsfonts}
\usepackage{amssymb}
\usepackage{color}
\newcommand{\etal}{\textit{et al.}}
\newcommand{\Xstate}{\mbox{$X^1\Sigma^+$}}
\newcommand{\astate}{\mbox{$a^3\Sigma^+$}}

\begin{document}

\title{Feshbach Loss Spectroscopy in an Ultracold $^{23}$Na and $^{40}$K Mixture}
\author{Min-Jie Zhu$^{1,2,3}$}
\author{Huan Yang$^{1,2,3}$}
\author{Lan Liu$^{1,2,3}$}
\author{De-Chao Zhang$^{1,2,3}$}
\author{Ya-Xiong Liu$^{1,2,3}$}
\author{Jue Nan$^{1,2,3}$}
\author{Jun Rui$^{1,2,3}$}
\author{Bo Zhao$^{1,2,3}$}
\author{Jian-Wei Pan$^{1,2,3}$}
\affiliation{$^{1}$Shanghai Branch, National Laboratory for Physical Sciences at Microscale and
Department of Modern Physics, University of Science and Technology of China,
Hefei, Anhui 230026, China}
\affiliation{$^{2}$CAS Center for Excellence and Synergetic Innovation Center in Quantum
Information and Quantum Physics, University of Science and Technology of
China, Shanghai 201315, China}
\affiliation{$^{3}$CAS-Alibaba Quantum Computing Laboratory, Shanghai 201315, China}

\author{Eberhard Tiemann$^{4}$}
\affiliation{$^{4}$Institut f\"{u}r Quantenoptik, Leibniz Universit\"{a}t Hannover, 30167
Hannover, Germany}

\begin{abstract}
We perform Feshbach spectroscopy in an ultracold mixture of $^{23}$Na and $^{40}$K with different spin-state combinations. We have observed 24 new interspecies Feshbach resonances at magnetic field up to 350 G. A full coupled-channel calculation is performed to assign these resonances. Among them, 12 resonances are identified as \emph{d}-wave Feshbach resonances. These \emph{d}-wave Feshbach resonances are about 5 G systematically smaller than the predications based on previous model potential. Taking into account these new experimental results, we improve the  Born-Oppenheimer potentials between Na and K, and achieve good agreement between the theory and experiment for all the observed Feshbach resonances.
\end{abstract}

\maketitle

\section{Introduction}

Tunable Feshbach resonances are an important tool in ultracold atomic gases and have
been widely used in studying strongly interacting quantum gases and
associating Feshbach molecules~\cite{chin2010feshbach,giorgini2008theory,koehler2006}. For  ultracold atomic mixtures,
interspecies Feshbach resonances can be employed to create polar molecules~\cite{ni2008high,takekoshi2014ultracold,molony2014creation,guo2016creation}
and investigate few-body Efimov physics~\cite{braaten2006universality,kraemer2006evidence,pollack2009universality,zaccanti2009observation,gross2009observation,huckans2009three,berninger2011universality,wild2012measurements,roy2013test,pires2014observation,tung2014geometric}.  Feshbach spectroscopy
is also a high-precision method to determine the long range form of the ground state
Born-Oppenheimer potentials~\cite{samuelis2000cold,allard2002ground,pashov2005potentials,docenko2006coupling,knoop2011feshbach,viel2016feshbach,tiemann2009coupled}. Recently, interspecies Feshbach resonances in
different ultracold atomic mixtures have been experimentally investigated~\cite{koppinger2014production,repp2013observation,inouye2004observation,wille2008exploring,ferlaino2006feshbach,pilch2009observation,tiecke2010broad,papp2006observation}.
Among them the Feshbach resonance between $^{23}$Na and $^{40}$K  has attracted particular
attention~\cite{park2012quantum,wu2012ultracold,park2015ultracold,rui2017controlled}. In Ref.~\cite{park2012quantum}, 32 \emph{s}-wave and \emph{p}-wave Feshbach resonances have been observed for this Bose-Fermi mixture. Broad \emph{s}%
-wave Feshbach resonances have been used to prepare weakly bound Feshbach molecules~\cite{wu2012ultracold} and chemically stable polar
molecules~\cite{park2015ultracold}. The overlapping Feshbach resonances for different
spin-state combinations have been employed to study ultracold chemical reaction with weakly bound Feshbach molecules~\cite{Knoop2010,Incao2009,rui2017controlled}. These Feshbach resonances may also be employed to implement quantum simulation of the Kondo effect with ultracold molecules~\cite{bauer2013realizing}.

In this work, we report on an extensive experimental study of Feshbach loss
spectroscopy in an ultracold $^{23}$Na-$^{40}$K mixture with
spin-state combinations or magnetic fields different from the work by Park \etal~\cite{park2012quantum}, which allows the study of a different regime of hyperfine coupling. We have observed 24 interspecies Feshbach
resonances at magnetic fields up to 350 G. We perform a full coupled-channel
calculation based on the previous model potential \cite{gerdes2008ground,temelkov2015molecular} to identify these resonances. Among them, 8
\textit{s}-wave and 4 \textit{p}-wave resonances can be reproduced by the theory. However, the remaining 12 Feshbach resonances, which are later assigned as $d$-wave Feshbach resonances, are about 5 G systematically smaller than theoretical predications. The Born-Oppenheimer potentials of the ground states \Xstate~and \astate~ from Refs.~\cite{gerdes2008ground,temelkov2015molecular}
are then adjusted by taking into account these \emph{d}-wave resonances, and good agreement with the experimental results is obtained for all the observed Feshbach resonances. The improved Born-Oppenheimer potentials are important to determine the interspecies Feshbach resonances between $^{23}$Na and $^{39}$K and between $^{23}$Na and $^{41}$K.

In section \ref{exp}, we will first introduce the experimental apparatus and the
procedures to create the ultracold $^{23}$Na-$^{40}$K mixture. We then
discuss the Feshbach spectroscopy measurements and present the results. In section \ref{theory} a
full coupled-channel calculation is performed to analyze the results and conclusions are drawn in section \ref{conc}.

\begin{figure}[tbp]
\includegraphics[width=8cm]{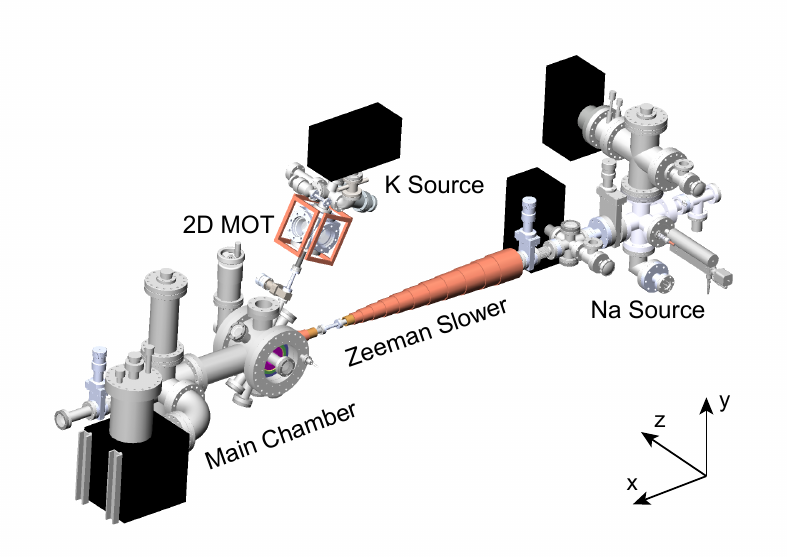}
\caption{Illustration of experimental setup. Zeeman slower and 2-dimensional magneto-optical trap (MOT) produce a slow flux of $^{23}$Na and $^{40}$K atoms, respectively. The atoms are captured in the main chamber by a two-species dark MOT, and subsequently transferred into a cloverleaf magnetic trap for evaporative cooling. In the final stage, the atoms are loaded into a crossed-beam dipole trap for further evaporative cooling to achieve ultracold Bose-Fermi mixture.}
\label{fig1}
\end{figure}

\section{Experiment}
\label{exp}
The experimental apparatus is depicted in Fig.~\ref{fig1}. We employ a Zeeman slower and
2D magnetic-optical-trap (MOT) to produce atom fluxes of $^{23}$Na and $^{40}$K, respectively. The atoms are then captured by a two-species dark MOT,
which can suppress the light-assisted interspecies collisions. Since the loading rate of
Na is much larger than that of K, we employ a two-species loading
sequence, i.e. loading K for about 20 s and loading Na in the last 2 s. In
this way, we obtain about $1\times 10^{9}$ Na and $1\times 10^{7}$ K
atoms. After the MOT loading stage, we perform high field Zeeman pumping to
prepare  Na in the $|2,2\rangle $ and K in the $|9/2,9/2\rangle $ state. The
bias field for optical pumping is about 140 G, and the pumping durations for
K and Na are 1 ms and 2 ms, respectively.

After the optical spin polarization, the atoms are captured by a cloverleaf
Ioffe-Pritchard magnetic trap with a bias magnetic field of about 140 G, a radial
magnetic field gradient of about 128 G/cm and an axial quadrature of about 79 G/cm$^{2}$. Further spin purification of Na is performed by applying a 1 s microwave
pulse coupling $|2,1\rangle \rightarrow |1,0\rangle $ and $|2,0\rangle
\rightarrow |1,1\rangle $ to eliminate the remaining atoms in $|2,1\rangle $
or $|2,0\rangle $ state. Subsequently, the magnetic trap is compressed by
reducing the bias magnetic field to about 1 G, and forced evaporative
cooling of Na is performed by sweeping the microwave frequency near 1.77 GHz
for 19 s. K is sympathetically cooled by elastic collisions with Na. To
suppress three-body losses, the magnetic trap is then decompressed by
adiabatically increasing the bias field to 12 G and reducing the radial
gradient to about 96 G/cm. Evaporative cooling continues in the decompressed trap
for another 3 s.

\begin{figure}[tbp]
\includegraphics[width=8cm]{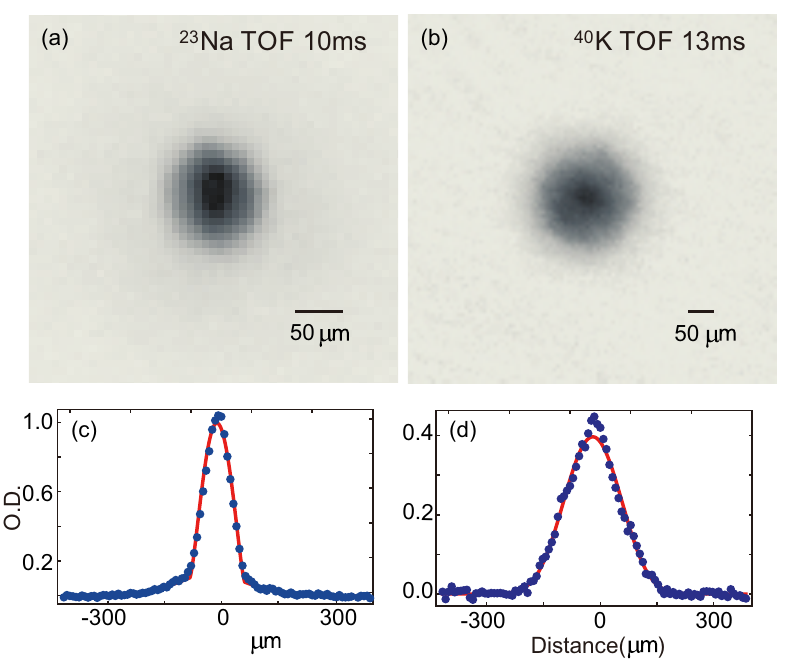}
\caption{(a) and (b) are the absorption images of the quantum degenerate Bose-Fermi mixture. (c) and (d) are center-sliced column density of the
bosonic and fermionic clouds. The solid lines in (c) and (d) are bimodel and Fermi-Dirac fitting curve for Na and K, respectively. }
\label{fig2}
\end{figure}

The atoms are transferred into a crossed-beam optical dipole trap (wavelength 1064nm),
consisting of a horizontal beam (power 5.5 W, waist 65 $\mu $m) and a
vertical beam (power 12.5 W, waist 123 $\mu $m). The optical dipole trap is
switched on in 30 ms and the magnetic trap is switched off suddenly. In the
optical dipole trap, Na atoms in $|2,2\rangle $ state are transferred to $|1,1\rangle $ by a Landau-Zener process to avoid significant three-body
losses between $|2,2\rangle $ and K atoms in $|9/2,9/2\rangle $ state.
Further evaporative cooling is done by lowering the power of the trapping
laser. By reducing the power of the horizontal beam and vertical beam to 4\%
and 18\% of their initial values, quantum degenerate Bose-Fermi
mixture of $^{23}$Na and $^{40}$K can be created. Absorption images of the samples are presented in Fig.~\ref{fig2}. We typically produce $1.2\times 10^{5}$ K atoms at a temperature of about $220$ nK coexisting with
$6\times 10^{4}$ Bose-condensed Na atoms. A Fermi-Dirac profile  fit (see Fig.~\ref{fig2}(d)) of the K
time of flight images gives $T\approx 0.5T_{F}$.

To perform Feshbach spectroscopy, we adiabatically increase the power of the
dipole trap by 3 times in 100 ms to better hold the atoms and increase
three-body collision rate. In this trap, the temperature of the atomic
mixture is about 1.5 $\mu $K and the lifetime of the ultracold mixture is
longer than 10 s, which is sufficient for the observation of Feshbach
resonances.

The atomic mixture is stable against spin exchanging collisions if one of
the species is in the absolute ground state and the other is in a low lying
hyperfine state. Therefore, we keep the Na in $|1,1\rangle $ state
and prepare K in different $|9/2,m_{f}\rangle $ Zeeman states. By applying a
50 ms radio-frequency (RF) Landau-Zener-sweep pulse at a bias field of 15 G,
the K atoms are transferred to a desired $m_{f}$ state with an efficiency higher than 99\%.

\begin{figure}[tbp]
\includegraphics[width=8cm]{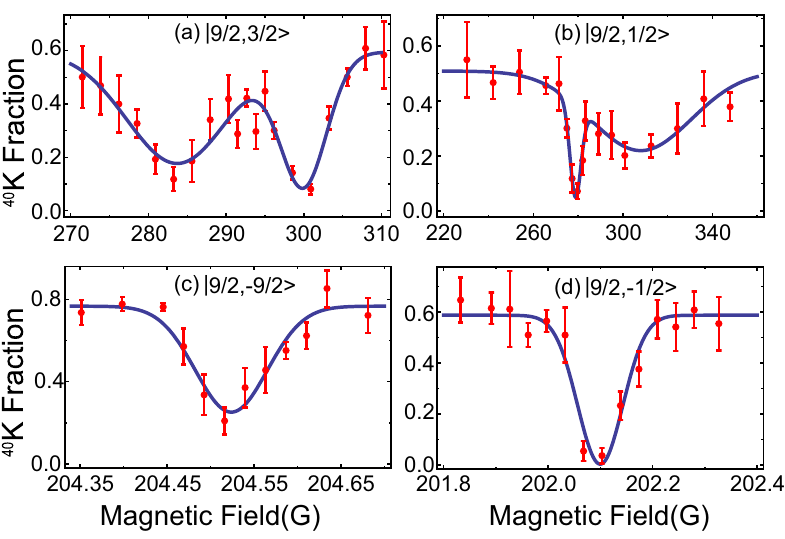}
\caption{The observed loss spectrum as a function the magnetic field. The solid lines are the Gaussian fitting curves. (a): A broad \emph{d}-wave Feshbach resonance near 283 G between Na in $|1,1\rangle$ and K in $|9/2,3/2\rangle$ states accompanied by a slightly narrower \emph{s}-wave resonance. (b): similar situation for the pair Na in $|1,1\rangle$ and K in $|9/2,1/2\rangle$ states. These \emph{d}-resonances show up by enhanced two-body inelastic collisions. (c) and (d): Narrow \emph{d}-wave Feshbach resonances near 200 G between Na in $|1,1\rangle$ and K in $|9/2,-9/2\rangle$ and $|9/2,-1/2\rangle$ states. These resonances are observed through 3-body losses. }
\label{fig3}
\end{figure}

The Feshbach magnetic field is created by the anti-bias coil of the
cloverleaf trap. The magnetic field is actively stabilized by a feedback
circuit and is calibrated by performing RF spectroscopy on the K $|9/2,-9/2\rangle \rightarrow |9/2,-7/2\rangle $ transition leading to an accuracy of the field calibration of about 10 mG. The stability of the magnetic field is about 20 mG. The atoms are held in the high
magnetic field for a few hundred milliseconds and then imaged at zero field.
Feshbach resonances are determined by the observation of enhanced loss at a
specific value of magnetic field. We typically prepare 4-5 times more Na than K
atoms and observe the loss of K atoms. Additional measurements are performed on pure Na
or K atoms to make sure that the observed enhanced loss is caused by an
interspecies Feshbach resonance. The loss profiles are phenomenally
fitted to Gaussian functions, from which the resonance positions $B_{\rm{exp}}$ and widths $\Delta_{\rm{exp}}$ are
obtained. The results are
summarized in Table~\ref{Table1}. We observed 24 Feshbach
resonances including 4 \emph{p}-wave resonances with their expected splitting from the effective spin-spin interaction, in total 28 observations. A full coupled-channel calculation (see section \ref{theory}) is performed to assign these resonances.

Four examples of experimental recordings are shown in Fig~\ref{fig3}. In the upper part fairly broad structures appear, where the losses are mainly caused by two-body inelastic processes as discussed in section \ref{theory}, whereas the lower part gives examples for narrow features later assigned to \emph{d}-wave resonances. Note the  very different magnetic field scales in the figure. For narrow resonances with a width lower than 0.4 G the peak could be determined with an accuracy of about 0.1 G due to calibration uncertainty and stability of the magnetic field, but for the very broad features we estimate that the accuracy is limited to about 20 \% or more of their widths due to the low signal-to-noise ratio at their wings (see Fig. \ref{fig3} (a) and (b)).

\begin{table}[h]
\begin{tabular}{c|c|c|c|c}
\hline\hline
$^{40}$K $|F,m_F\rangle$ & $B_{\mathrm{exp}}$(G) & $\Delta_{\mathrm{exp}}$(G)& $B_{\mathrm{cc}}$(G) & $l$ \\ \hline
$|9/2,-1/2\rangle$ & 146.7 & 0.3 &146.94  & $s$ \\ \hline
& 165.3 & 2.3 &165.72  & $s$ \\ \hline
& 233.0 & 18.3 & 238.1 & $s$ \\ \hline
& 18.81 & 0.11 &18.85  & $p$ \\ \hline
&  19.15 & 0.15 & 19.05 & $p$ \\ \hline
& 58.32 & 0.07 &58.36  & $p$ \\ \hline
& 59.10 & 0.13 & 58.86 & $p$ \\ \hline
$|9/2, 1/2\rangle$ & 190.5 & 0.2 & 191.01 & $s$ \\ \hline
& 218.4 & 1.4 &219.02  & $s$ \\ \hline
& 308.1 & 31.9 & 327.7 & $s$ \\ \hline
& 35.17 & 0.11 &35.29  & $p$ \\ \hline
&  35.83 & 0.19 &  35.74& $p$ \\ \hline
& 100.36 & 0.23 & 100.26 & $p$ \\ \hline
&  101.31 & 0.42 & 100.93 & $p$ \\ \hline
$|9/2, 3/2\rangle$ & 256.6 & 1.1 & 257.29 & $s$ \\ \hline
& 299.9 & 4.2 & 301.5 & $s$ \\ \hline\hline
$|9/2, -9/2\rangle$ & 204.52 & 0.06 & 204.47 & $d$ \\ \hline
& 279.8 & 0.1 & 280.06 & $d$ \\ \hline
$|9/2, -7/2\rangle$ & 202.68 & 0.07 &202.61  & $d$ \\ \hline
& 276.3 & 0.3 &276.40  & $d$ \\ \hline
$|9/2, -5/2\rangle$ & 201.66 & 0.04 & 201.57 & $d$ \\ \hline
& 274.6 & 0.3 &274.62  & $d$ \\ \hline
$|9/2, -3/2\rangle$ & 201.44 & 0.10 & 201.34 & $d$ \\ \hline
& 274.8 & 1.2 & 274.62 & $d$ \\ \hline
$|9/2, -1/2\rangle$ & 202.10 & 0.06 & 201.95 & $d$ \\ \hline
& 276.2 & 13.4 &276.28  & $d$ \\ \hline
$|9/2, 1/2\rangle$ & 278.8 & 3.47 &279.54  & $d$ \\ \hline
$|9/2, 3/2\rangle$ & 283.7 & 9.1 & 283.73 & $d$ \\ \hline\hline
\end{tabular}
\caption{Feshbach resonances between $^{23}$Na in $|1,1\rangle$ and $^{40}$K in different internal states. $B_{\mathrm{exp}}$ and $\Delta_{\mathrm{exp}}$ are the positions and widths of the resonances determined from the loss spectrum by phenomenal Gaussian fitting. $B_{\mathrm{cc}}$ is the theoretical result obtained by the coupled-channel calculation. $l$ denotes the resonance type.
 }
\label{Table1}
\end{table}

\section{Modeling the observations}
\label {theory}
For the description of the scattering resonances we set up a Hamiltonian for the nuclear motion of the two atoms, i.e. pair rotation and radial motion with molecular potentials, and their hyperfine interaction. The electronic space contains the singlet and triplet ground states with their electronic assignments \Xstate~and \astate, respectively. The appropriate Hamiltonian has already been given in many papers~\cite{hutson2008avoided,falke2008potassium,temelkov2015molecular,knoop2011feshbach,laue2002magnetic,mies1996estimating} and thus not reproduced here. We start the present modeling using the hyperfine and Zeeman parameters compiled by Arimondo \etal \ in 1977~\cite{arimondo1977experimental} and potential functions for both electronic states reported by Temelkov \etal \cite{temelkov2015molecular}. This earlier evaluation combined high resolution molecular beam spectroscopy \cite{temelkov2015molecular} and Feshbach spectroscopy by Park \etal~\cite{park2012quantum}. Thus we expected that the predictions directly from this model should be fairly good. But it turned out that significant deviations appeared and some of the observations could not be unambiguously assigned. This results mainly from the fact that the Feshbach resonances observed in \cite{park2012quantum} are related to bound states, which predominantly have triplet character, and thus the scattering length of the singlet channel is not yet well determined. The new measurements contain resonances related with bound states at zero field showing expectation values of the total electron spin of 0.76 for the later assigned \emph{d}-wave resonances compared to 0.98 in the former cases. The predictions of \emph{d}-wave resonances are systematically at higher field values than observed features.

\begin{figure}[tbp]
\includegraphics[width=10cm]{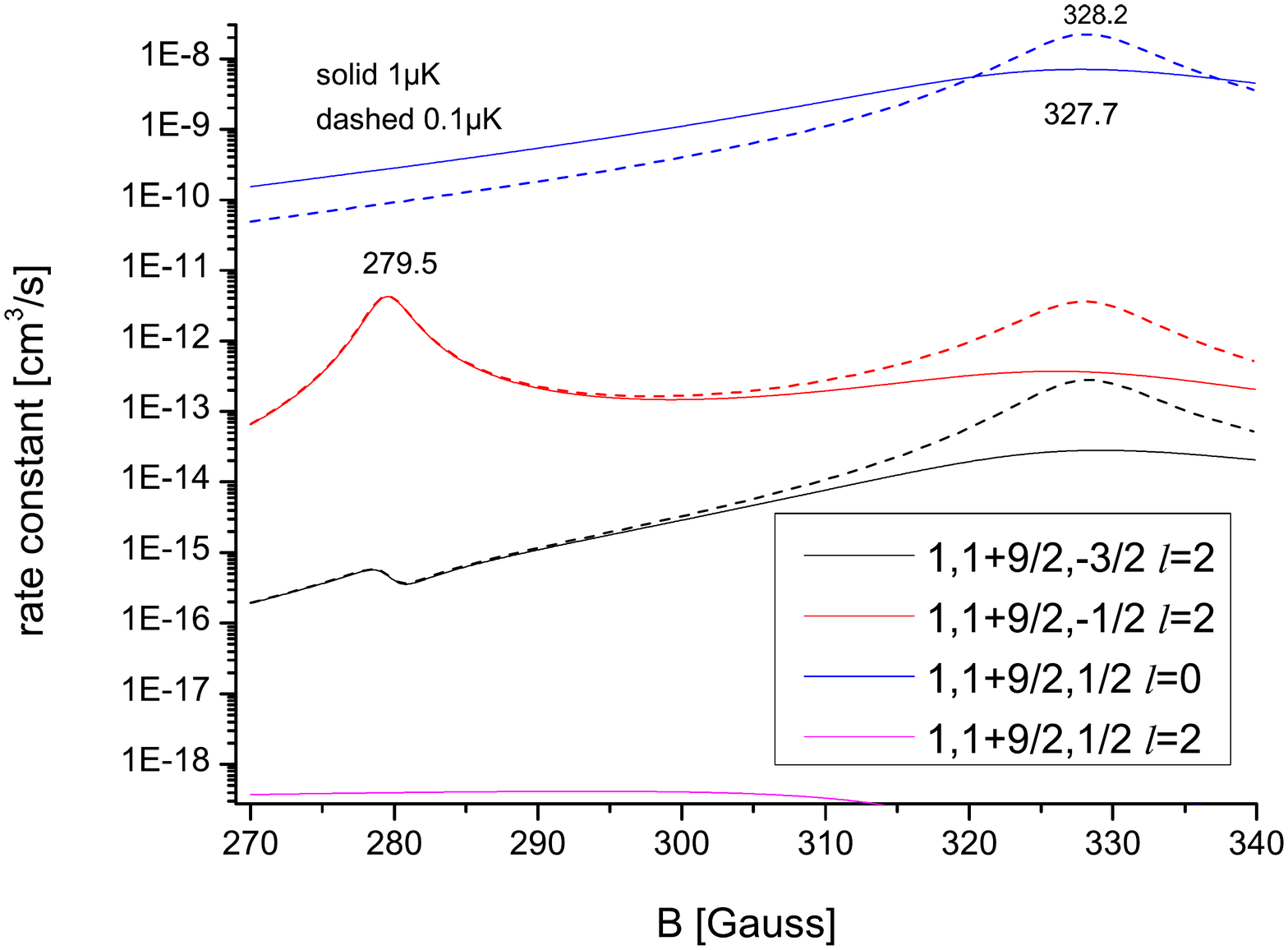}
\caption{Calculated Feshbach resonances for two different kinetic energies between Na in $|1,1\rangle$ and K in $|9/2,1/2\rangle$ states. The feature at 278.5 G is a \emph{d}-resonance and that at 327.7 G a \emph{s}-resonance. The elastic and inelastic channels are labeled by atom pair quantum numbers in the right box. }
\label{fig4}
\end{figure}

Thus we started the evaluation by using the new measurements, which could fairly reliably assigned as \emph{s}-wave resonances and fitted the potentials keeping the observations from Park \etal \cite{park2012quantum} also in the fit. Then new predictions for \emph{s}-wave and \emph{p}-wave resonances were made and compared to the observations. By this kind of iteration process, 16 observations in Table. \ref{Table1} could unambiguously be assigned and we could exclude that the remaining 12 observed resonances (lower part of the table) would be \emph{s} or \emph{p} ones. The purity of the atom pair state preparation is better than 99\%, thus the remaining observations are not belonging with high probability to undesired atom pair collisions. Thus we searched for locations of \emph{d}-wave resonances with the best model potentials obtained from the  evaluation up to this point. The narrow feature at 279.8 G for the pair $|1,1\rangle$ + $|9/2,-9/2\rangle$ was fairly close to a predicted \emph{d}-resonance and relying on this assignment we calculated a very narrow resonance around 204 G, which was soon found (see Fig. \ref{fig3} (c) and confirmed the successful start of identifying \emph{d}-resonances. With these first data we made improvements on the potential scheme of the \Xstate~and \astate~states.

For identifying the additional features close to \emph{s}-wave resonances as shown in the upper part of Fig. \ref{fig3}, we found, that the resonance did not appear in the elastic but in the inelastic channel, an example is shown in Fig. \ref{fig4}. The experiment works with an ensemble of 1.5 $\mu$K. Thus we show two cases of kinetic energy for the collision process, namely 0.1 and 1.0 $\mu$K within the thermal distribution. One sees the broad features of the \emph{s}-resonance at the high field side, appearing in the elastic channel and also in the inelastic cases $|1,1\rangle$ + $|9/2,-1/2\rangle$ and $|1,1\rangle$ + $|9/2,-3/2\rangle$. These outgoing waves lead directly to losses without any additional third body partner because of the significant kinetic energy in the order of 3.5 mK for the dominant part  $|1,1\rangle$ + $|9/2,-1/2\rangle$. But at the low field side the resonance only appears in the inelastic channels and mainly in $|1,1\rangle$ + $|9/2,-1/2\rangle$. One cannot identify any trace of it in the elastic contribution. The two-body inelastic rate is not large compared to the elastic contribution, but the loss signal is significant, because it is a two-body process. The similarity of the calculated inelastic profile compared to the observation in Fig.\ref{fig3}(a) is striking, considering an average over temperature. In the same way the other broad features not yet assigned were analyzed and they are all \emph{d}-resonances with a strong inelastic contribution.

The final evaluation was performed by fitting the resonances to elastic or inelastic peaks of  two-body collisions. Also the data from \cite{park2012quantum} were taken into account to obtain the most reliable potentials for both electronic states, the effective spin-spin interaction and the hyperfine coupling as derived in \cite{temelkov2015molecular} were used unaltered. The derived analytical potential functions can be found in the supplementary material \cite{suppl}.

\section{Conclusion}
\label{conc}
We present 28 new Feshbach resonances between $^{23}$Na and $^{40}$K and perform a coupled channel fit of all known Feshbach resonances (in total 60) to obtain improved potential functions for both \Xstate~and \astate~states. The fit shows that the broad \emph{s}-wave resonances, like those given in Fig. \ref{fig4}, deviate significantly in their calculated and observed peak positions, namely in the order of their width. The reason is not clear; we checked that the close proximity between \emph{d}- and \emph{s}-resonances as seen in Fig. \ref{fig3} (a) and (b) is not related with this effect, because calculating the \emph{s}-resonance as a pure \emph{s}-wave or allowing the mixing of \emph{s}- and \emph{d}-partial waves shifts the resonance position by less than 0.2 G. Fig. \ref{fig4} gives calculations for two different kinetic energies, besides a difference between the \emph{s}-resonances of 0.5 G, only the peak height has changed. Thus thermal averaging will not result in a shift of about 20 G, as desired looking to results in Table. \ref{Table1} for this broad resonance of 32 G. We estimated the experimental accuracy of this case to be 10 G. Park \etal~\cite{park2012quantum} found similar discrepancies for broad resonances and the same group reported in \cite{wu2012ultracold} on measurements of binding energies of the corresponding Feshbach molecules which will result in a more reliable determination of the molecular state responsible for the Feshbach resonance.

\begin{table} [h]
\begin{tabular}{c|c|c|c}
\hline \hline
isotope & \Xstate & \astate  & reference   \\ \hline
$^{23}$Na + $^{39}$K & 324 (10) & -83.9 (10)& Temelkov~\cite{temelkov2015molecular}  \\
 & 255 & -84& Viel~\cite{viel2016feshbach}  \\
 & 331.8 (20) & -83.97 (50)& present  \\  \hline
$^{23}$Na + $^{40}$K & 66.4 (10) &-823 (5) & Temelkov \\
& 63 &-838 & Viel \\
 & 66.7 (3) &-824.7(30) & present \\ \hline
$^{23}$Na + $^{41}$K & 2.88 (40)& 267.2 (10) & Temelkov \\
 & -3.65& 267& Viel \\
 & 3.39 (20)& 267.05 (50) & present \\
\hline \hline
\end{tabular}
\caption{
Derived scattering lengths in atomic unit ($a_0=0.529\times10^{-10}m$) for the different isotopes of Na$+$K collisions.}
\label{Table2}
\end{table}

The Feshbach data from \cite{park2012quantum} were evaluated applying coupled channel calculations by Temelkov \etal \cite{temelkov2015molecular} and by Viel and Simoni \cite{viel2016feshbach}. These authors give their compact result on the long range behavior of \Xstate~and \astate~states through a table of scattering lengths for the three isotope combinations, possible from natural isotopes of potassium. We use the potentials derived in this work for calculations of these quantities and compare the results in Table. \ref{Table2} with those published. Because the experimental data applied in the former evaluations are strongly related to the triplet state, the agreement for the triplet scattering lengths is very satisfying, but as expected differences show up for the values of the singlet state. Here the new values agree better with those from Temelkov~\cite{temelkov2015molecular} due to the inclusion of the molecular beam data, which are also related to the singlet state. We should mention, that the \emph{p}-wave resonances are not as good represented in the work by Viel and Simoni~\cite{viel2016feshbach} as in the others, because they did not extend the effective spin-spin interaction by the second order spin-orbit contribution. This is essential as already shown by Temelkov \etal \cite{temelkov2015molecular}.

Future experimental research includes the study of few-body physics and cold molecules employing these Feshbach resonances. 
The observed \emph{d}-wave Feshbach resonances offer new opportunity to study Efimov physics and cold molecules with high partial wave resonances. It is possible to form \emph{d}-wave Feshbach molecules by magnetic field sweeping association. Therefore, the overlapping \emph{d}-wave Feshbach resonances for different spin-state combinations may be employed to study the ultracold chemistry with high partial wave molecules, similar to the overlapping \emph{s}-wave Feshbach resonances~\cite{Incao2009,Knoop2010,rui2017controlled}.

\section{Acknowledgement}
We acknowledge support from the National
Natural Science Foundation of China (under Grant No.11521063, 11274292), the
National Fundamental Research Program (under Grant No. 2013CB336800), the
Chinese Academy of Sciences, and Shanghai Sailing Program (under Grant No. 17YF1420900).

\end{document}